\ifdefined\WORDCOUNT
\documentclass[superscriptaddress,twocolumn,preprintnumbers,amsmath,amssymb,prl,nofootinbib,groupedaddress]{revtex4-1}
\else
\documentclass[superscriptaddress,twocolumn,showpacs,preprintnumbers,amsmath,amssymb,prl,groupedaddress]{revtex4-1}
\fi

\usepackage{amssymb}
\usepackage{epsfig,amsmath,graphics}
\usepackage{epstopdf}
\usepackage{graphicx}
\usepackage{color}

\newcommand{\ket}[1]{\ensuremath{\left|#1\right>}}

\ifdefined\WORDCOUNT
    \newcommand{\HideEq}[1]{}
\else
    \newcommand{\HideEq}[1]{#1}
\fi

\def\be{\begin{equation}}
\def\ee{\end{equation}}

\begin{document}

\title{Effective inertial frame in an atom interferometric test of the equivalence principle}

\def\stanfordAffiliation{Department of Physics, Stanford University, Stanford, California 94305}

\author{Chris Overstreet}
\author{Peter Asenbaum}
\author{Tim Kovachy}
\author{Remy Notermans}
\author{Jason M. Hogan}
\author{Mark A. Kasevich}
\affiliation{\stanfordAffiliation}
\email{kasevich@stanford.edu}

\date{\today}

\begin{abstract}
In an ideal test of the equivalence principle, the test masses fall in a common inertial frame.  A real experiment is affected by gravity gradients, which introduce systematic errors by coupling to initial kinematic differences between the test masses. We demonstrate a method that reduces the sensitivity of a dual-species atom interferometer to initial kinematics by using a frequency shift of the mirror pulse to create an effective inertial frame for both atomic species. This suppresses the gravity-gradient-induced dependence of the differential phase on initial kinematic differences by a factor of 100 and enables a precise measurement of these differences. We realize a relative precision of $\Delta g / g \approx 6 \times 10^{-11}$ per shot, which improves on the best previous result for a dual-species atom interferometer by more than three orders of magnitude. By suppressing gravity gradient systematic errors to below one part in $10^{13}$, these results pave the way for an atomic test of the equivalence principle at an accuracy comparable with state-of-the-art classical tests. 
      
\end{abstract}


\ifdefined\WORDCOUNT

\else
   \maketitle
\fi

The equivalence principle lies at the heart of general relativity, and efforts to test its validity with increasing precision for a variety of test objects are at the forefront of experimental physics \cite{Williams2004, Schlamminger2008, Fray2004, Bonnin2013, Schlippert2014,Tarallo2014, Zhou2015, Barrett2016, Rosi2017b, Berge2015, Hogan2009, Altschul2015,Hartwig2015,Williams2016}.  Many of these experiments probe the weak equivalence principle (WEP), which stipulates the universality of free fall \cite{Carroll2004a}.  In addition to testing a fundamental aspect of general relativity, WEP tests can be used to search for new interactions and for dark matter \cite{Dimopoulos2008a,Graham2016}.  

All WEP tests operate under the same general principle--they compare the gravitational accelerations of two test masses of different composition.  In an ideal thought experiment, this comparison would occur in a uniform gravitational field, making the measurement insensitive to the initial kinematics of the test masses.  However, in realistic experimental setups, gravity gradients are present.  Gravity gradients cause the measured acceleration of a given test mass to vary linearly as a function of its initial position and velocity.  As a consequence, mismatches in the initial kinematics of the test masses can appear as a spurious WEP violation if not characterized to the necessary accuracy.  This coupling of initial kinematics to gravity gradients is a leading systematic error in WEP tests based on atom interferometry \cite{Hogan2009,Nobili2016}.

It is relevant to consider the ramifications of this effect for Earth's gravity gradient, which is approximately $T_{zz} = 3 \times 10^{-7} g/\text{m}$ in the vertical direction.  To lowest order, the differential acceleration that the gravity gradient induces between the test masses A and B is $g_\text{A} - g_\text{B} = T_{zz} [\Delta z + \Delta v\,T] \equiv T_{zz} \Delta \bar{z}$, where $\Delta z = z_A - z_B$, $\Delta v = v_A - v_B$, $g_i$, $z_i$, and $v_i$ are the respective gravitational acceleration, initial position, and initial velocity of test mass $i \in \{A,B\}$, and $T$ is the time interval over which the acceleration measurement occurs.  This implies, for example, that an equivalence principle test with relative accuracy $2(g_\text{A} - g_\text{B})/(g_\text{A} + g_\text{B}) = (g_\text{A} - g_\text{B})/g \approx 10^{-14}$ requires relative displacements arising from initial kinematics to be controlled at the level of 30~nm.

In this work, we experimentally demonstrate a method to make a dual-species atom interferometric WEP test \cite{Fray2004, Bonnin2013, Schlippert2014,Tarallo2014, Zhou2015, Barrett2016} insensitive to initial kinematics.  Following the proposal of Roura \cite{Roura2017}, the optical frequency is shifted for the mirror sequence of a light-pulse Mach-Zehnder atom interferometer \cite{Borde1989, Kasevich1991}, producing a phase shift proportional to the average vertical displacement $\Delta \bar{z}$ during the interferometer \cite{Biedermann2015}.  An appropriate choice of this frequency shift counteracts the corresponding phase shift from the gravity gradient \cite{Roura2017}, creating an effective inertial frame.  Although the interferometer trajectories remain perturbed by the gravity gradient as a function of initial position and velocity, the interferometer phase becomes insensitive to these perturbations. We refer to this method as frequency shift gravity gradient compensation (FSGG compensation). Using FSGG compensation in a long duration/large momentum transfer (long $T$/LMT) dual species interferometer with $^{85}$Rb and $^{87}$Rb, we reduce the sensitivity to initial kinematic mismatches to less than $1\%$ of its original value.  Moreover, we introduce a technique to determine the correct frequency shift without needing to independently measure or calculate the gravity gradient.  An analogous method to FSGG compensation is not currently known for classical free-fall WEP tests.

\begin{figure}[h!]
	\begin{center}
		\includegraphics[width=\linewidth]{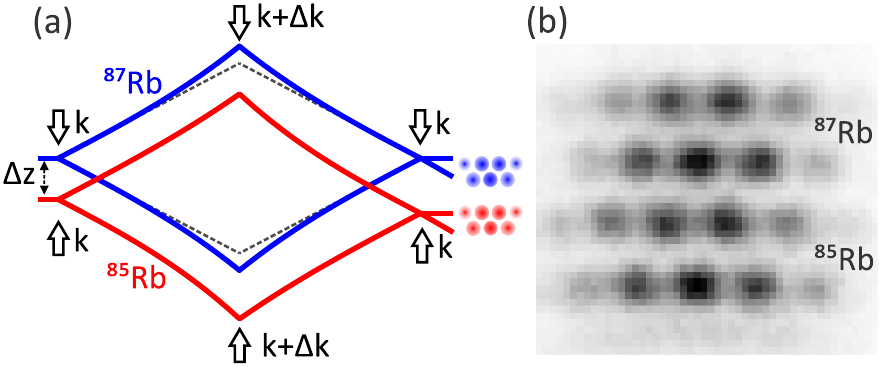}
        \caption{Implementation of the FSGG compensation scheme. (a) The interferometers in a dual-species differential accelerometer are separated by an initial displacement $\Delta z$.  Due to the gravity gradient $T_{zz}$, the interferometers experience a differential acceleration $T_{zz} \Delta z$ and a differential phase shift $k T_{zz} \Delta z T^2$. To perform FSGG compensation, the effective wavevector of the interferometer mirror pulses is changed by $\Delta k$, which adds a differential phase shift $-2 \Delta k\Delta z$. For $\Delta k/k = \frac{1}{2} T_{zz} T^2$, the differential phase becomes insensitive to the initial displacement $\Delta z$. Alternatively, a scan of $\Delta k$ provides information about $\Delta z$ \cite{Biedermann2015}. (b) Raw fluorescence image of the dual-species differential accelerometer operating at an LMT order of $10 \hbar k$ with initially overlapped clouds and $\Delta f = 343~\text{MHz}$, using phase shear readout to determine the differential phase.}
		\label{fig:figure1}
	\end{center}
\end{figure}

The core features of the experimental apparatus have been described in previous work \cite{Dickerson2013,Sugarbaker2013, Kovachy2015, Kovachy2015a, Asenbaum2017}.  Some modifications to the atom source have been made in order to generate an ultracold dual species cloud (earlier experiments used only $^{87}$Rb).  Approximately $4 \times 10^{9}$ $^{87}$Rb atoms and $3 \times 10^8$ $^{85}$Rb atoms are loaded from a 2D-MOT into a 3D-MOT.  Subsequently, forced microwave evaporation is performed on the $^{87}$Rb atoms in a quadrupole and then a time-orbiting potential (TOP) trap.  The $^{85}$Rb atoms are sympathetically cooled.  During evaporation, the $^{87}$Rb atoms are in the $\ket{F=2, m_F=2}$ state and the $^{85}$Rb atoms are in the $\ket{F=3, m_F=3}$ state.  Following a magnetic lensing sequence to collimate the atom clouds \cite{Kovachy2015}, an optical lattice launches the atoms upward into a 10~m fountain.  After the launch, an optical dipole lens provides further collimation in the transverse dimensions \cite{Asenbaum2017}, and the atoms are prepared in Zeeman insensitive hyperfine sublevels by a sequence of microwave pulses.  Residual atoms that are not transferred by the state preparation pulses are removed by momentum transfer from a resonant light pulse.  At the time of detection, the atom clouds have expanded to a radial size of approximately 1~cm.

Following the work described in \cite{Kovachy2015a,Asenbaum2017}, the LMT beam splitter and mirror sequences for the interferometer use absolute-AC-Stark-shift-compensated sequential two-photon Bragg pulses.  The Bragg pulses simultaneously address the $^{85}$Rb and $^{87}$Rb atoms so that phase shifts from optical path length fluctuations (e.g., due to vibrations) cancel as a common mode in the differential measurement.  Typical experimental parameters are $10 \hbar k$ momentum splitting between the interferometer arms and $T = 900~\text{ms}$ pulse spacing ($k$ denotes the wave number of the Bragg lasers). In most of the experimental runs, the beam splitters operate in a symmetric or near-symmetric configuration \cite{Ahlers2016}.  For instance, for a $12 \hbar k$ beam splitter, the lower interferometer path receives a $6 \hbar k$ downward momentum kick and the upper interferometer path receives a $6 \hbar k$ upward kick.  For a $10 \hbar k$ beam splitter, the lower path receives a $6 \hbar k$ downward kick and the upper path receives a $4 \hbar k$ upward kick.  Upward- and downward-kicking Bragg pulses occur sequentially and are interleaved.

Figure \ref{fig:figure1}(a) illustrates the dual species, FSGG compensated interferometer.  The Bragg lasers nominally have frequency $f$.  For all the LMT pulses that make up the interferometer mirror sequence, the laser frequency is shifted by an amount $\Delta f$. In a uniform gravity gradient, $\Delta f/f = \Delta k/k = T_{zz} T^2/2$ results in perfect compensation.  As the gradient in the 10 m fountain changes substantially as a function of height \cite{Asenbaum2017}, the ideal $\Delta f$ involves a weighted average of the gravity gradients experienced by the atoms at different heights \cite{Roura2017}.  A CCD camera records fluorescence images of the interferometer output ports for both species.  Because the $^{85}$Rb and $^{87}$Rb clouds overlap to within the cloud size, we implement a staggered imaging sequence.  First, near-resonant light for only one species is pulsed on, stopping the atoms of that species in place.  The atoms of the other species are allowed to fall for an additional 0.9~ms before being stopped so that the output ports of the second species are imaged resolvably below the output ports of the first species on the CCD [see Fig. \ref{fig:figure1}(b)].  Horizontal spatial fringes are put across the output ports by tilting the angle of the retroreflection mirror for the final beam splitter sequence (phase shear readout) \cite{Sugarbaker2013, Asenbaum2017}.  Comparing the phase of these fringes for $^{85}$Rb and $^{87}$Rb provides a differential acceleration measurement for each run of the experiment.

\begin{figure}[t!]
	\begin{center}
		\includegraphics[width=\linewidth]{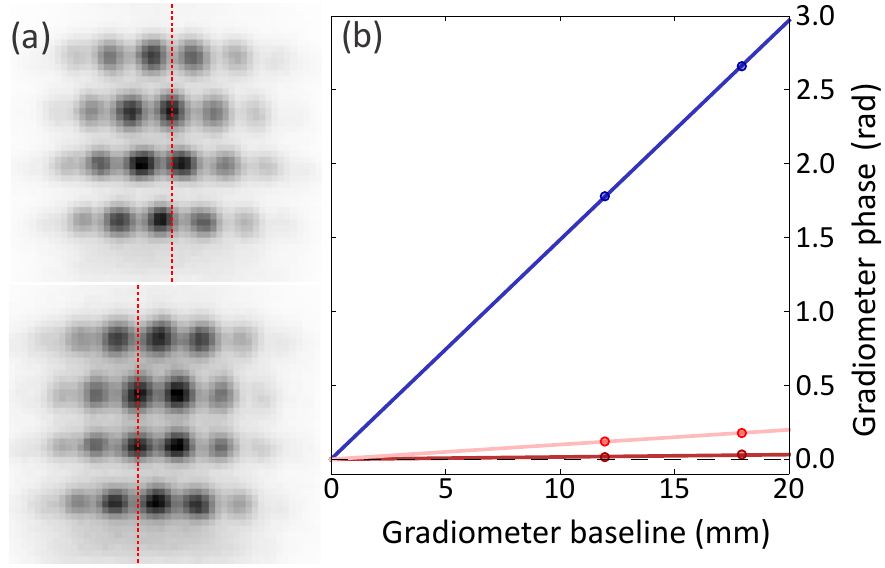}
			\caption{ (a) Raw data from $^{87}$Rb-only ($10 \hbar k$ momentum splitting, $T=900~\text{ms}$) gravity gradiometer used to determine the optimal frequency shift $\Delta f$.  The upper and lower pairs of output ports correspond to the two vertically displaced interferometers.  The two interferometers use opposite input ports, which would give them a differential phase of $\pi$ in the absence of any gravity gradients. Upper image:  without FSGG compensation.  Lower image: with FSGG compensation.  Without FSGG compensation, the differential phase is visibly shifted away from $\pi$.  With FSGG compensation, the differential phase is $\pi$, illustrating the cancellation of the gravity gradient phase shift.  (b)  Gradiometer phase vs. baseline for $\Delta f = 0~\text{MHz}$ (blue points), $\Delta f = 320~\text{MHz}$ (light red points), and $\Delta f = 340~\text{MHz}$ (dark red points).  Error bars are smaller than the points.}
		\label{fig:figure3}
	\end{center}
\end{figure}

To extract the differential phase from a fluorescence image, we bin each interferometer port vertically and compute the asymmetry $A(x) \equiv (P_1(x) - P_2(x)) / (P_1(x) + P_2(x))$ for each interferometer, where $P_i(x)$ is the number of counts in port $i$ as a function of horizontal position $x$.  Each interferometer asymmetry is then filtered and fit to a sinusoid.  For the data presented in this work, the single-shot differential phase uncertainty is typically $\sim 40$ mrad.

An accurate, a priori determination of the compensation frequency shift $\Delta f$ would require a sequence of many gravity gradient measurements with high spatial resolution.  It is more convenient to determine $\Delta f$ empirically by directly minimizing the sensitivity of the interferometer to initial kinematics.  Initial kinematic mismatches enter both the gravity-gradient-induced and FSGG-compensation phase shifts via the quantity $\Delta \bar{z}$ \cite{Roura2017}.  Since the interferometer is intrinsically velocity-selective, it is most convenient to vary $\Delta \bar{z}$ by adjusting $\Delta z$.  To optimize $\Delta f$, we used a $^{87}$Rb-only gravity gradiometer consisting of two simultaneous, vertically-displaced interferometers (see \cite{Asenbaum2017} for a description of the gradiometer sequence).  The optimal $\Delta f$ is determined by minimizing the displacement dependence of the differential phase shift between these interferometers (see Fig. \ref{fig:figure3}). This technique is operationally similar to methods for finding magic wavelengths in precision spectroscopy \cite{Boyd2007, Ye2008} and assumes that the gravity gradient is temporally stable.

\begin{figure}[t!]
	\begin{center}
		\includegraphics[width=\linewidth]{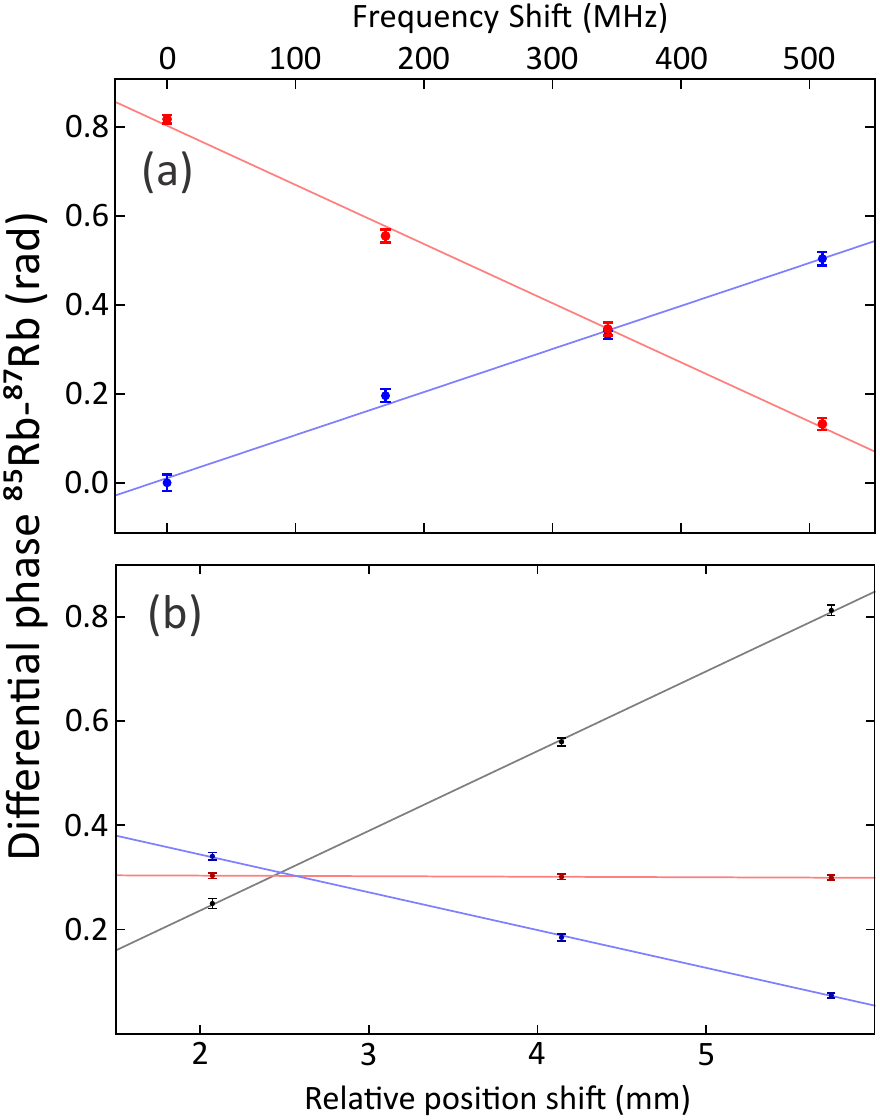}
				\caption{Dependence of differential phase on mirror pulse frequency shift $\Delta f$ and initial separation $\Delta z$. (a) Differential phase as a function of $\Delta f$ for two initial separations differing by $5.5~\text{mm}$. The slope of each linear fit is used to determine the quantities $\Delta \bar{z} = -3.21 \pm 0.07~\text{mm}$ (red points) and $\Delta \bar{z} = 2.3\pm 0.1~\text{mm}$ (blue points).  Each point is the average of $\sim 20$ experimental shots.  For $\Delta f = 345\pm 11~\text{MHz}$, the differential phase at the two separations is equal and therefore insensitive to the gravity gradient. (b) Differential phase as a function of relative position shift for $\Delta f = 0~\text{MHz}$ (black points, $\sim 30$ shots per point), $\Delta f = 510~\text{MHz}$ (blue points, $\sim 100$ shots per point), and $\Delta f = 343~\text{MHz}$ (red points, $\sim 100$ shots per point). The intersection point of all three lines provides the relative position shift required to set $\Delta \bar{z} = 0$, $2.47\pm 0.04~\text{mm}$. The slope of the red linear fit is $(0.7\pm 1.3)\%$ of the slope of the black linear fit, demonstrating the reduction of differential phase sensitivity to initial kinematic mismatches. All differential phases are referenced to the differential phase at $\Delta f = 0~\text{MHz}$ and $\Delta \bar{z} = 2.3 \pm 0.1~\text{mm}$.}
		\label{fig:figure2}
	\end{center}
\end{figure}

As a confirmation, we obtained the same value for $\Delta f$ using the dual species interferometer.  Specifically, for multiple values of $\Delta f$, we measure the variation of the differential phase when the initial displacement $\Delta z$ between the $^{85}$Rb and $^{87}$Rb clouds is shifted by 5.5~mm. The optimal $\Delta f$ is that for which the differential phases are equal, as shown in Fig. \ref{fig:figure2}(a). Note that this procedure does not suppress a possible EP-violating differential phase shift.  Figure \ref{fig:figure2}(b) shows that FSGG compensation cancels the $T_{zz} \Delta \bar{z}$ phase shift to $0.7 \pm 1.3\%$ of its uncompensated value, limited by statistical uncertainty.  

The data shown in Fig. \ref{fig:figure2} require the ability to independently adjust the positions of the $^{85}$Rb and $^{87}$Rb clouds at the start of the interferometer.  These adjustments are accomplished with the aid of two-photon Raman transitions \cite{Kasevich1991b, Kasevich1991}.  Unlike the Bragg transitions used in the interferometer, Raman transitions change the atomic hyperfine state.  Since $^{85}$Rb and $^{87}$Rb have significantly different ground state hyperfine splittings (3~GHz vs. 6.8~GHz) \cite{Steck85,Steck87}, Raman transitions can transfer momentum to one species while being far off resonance from the other.  An initial velocity-selection Bragg pulse occurs 130~ms before the first interferometer beam splitter.  Next, a Raman pulse delivers a $2 \hbar k$ momentum kick to the $^{87}$Rb atoms.  A corresponding Raman pulse for the $^{85}$Rb atoms can either be applied immediately following the $^{87}$Rb Raman pulse or after a delay of up to 120~ms, during which the two clouds move relative to each other.  During this delay time, an additional momentum offset between the $^{85}$Rb and $^{87}$Rb atoms can be achieved by further accelerating the $^{87}$Rb atoms with Bragg pulses.  We typically use a total momentum offset of $8 \hbar k$.  At the end of the delay time, a Bragg pulse deceleration sequence reverses these auxiliary momentum kicks. Varying the delay time allows for the tuning of the relative position shift between the two species.  With this technique, we can tune $\Delta \bar{z}$ to zero with an accuracy of $40~\mu\text{m}$, limited by the uncertainty of the slopes in Fig. \ref{fig:figure2}(b).  Combined with the suppression of initial kinematic sensitivity provided by FSGG compensation, this reduces the relative differential phase shift associated with $T_{zz} \Delta \bar{z}$ to $8 \times 10^{-14}$.  Improved statistical resolution, which will be present during the integration over many shots for an equivalence principle test, should allow the accuracy to which $\Delta \bar{z}$ is tuned to zero to be improved by more than a factor of 10.  This would bring the systematic error from $T_{zz}\Delta \bar{z}$ to below $1 \times 10^{-14}$.     
  
\begin{figure}[t!]
	\begin{center}
		\includegraphics[width=\linewidth]{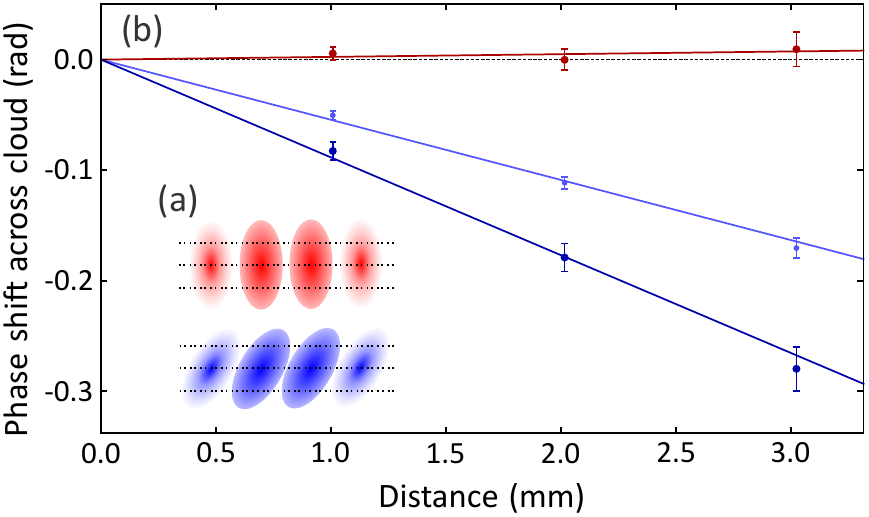}
				\caption{(a) Output ports of a $^{87}$Rb interferometer with (blue) and without (red) a phase shift in the vertical direction.  Each port is divided into four vertical bins.  (b) Phase shift as a function of vertical position at two LMT orders ($12 \hbar k$, light blue points; $20\hbar k$, dark blue and red points), with (red) and without (light blue, dark blue) FSGG compensation.  Distances are referenced to the top of each port.  The ratio of the slopes of the light blue and dark blue linear fits is $0.62 \pm 0.03$, consistent with the ratio of the LMT orders.  The slope of the red linear fit is $(-3 \pm 3)\%$ of the slope of the dark blue linear fit.  All phase shifts are referenced to the phase shift at the same cloud position of a $12\hbar k$ interferometer with FSGG compensation.}
		
		\label{fig:figure4}
	\end{center}
\end{figure}

Because the interferometer output ports have a finite spatial extent, the gravity gradient induces a phase shift across each output port in the vertical direction.  If the ports are vertically binned to extract the interferometer phase, averaging over the position-dependent phase shift reduces the contrast of the interferometer \cite{Roura2017}.  This effect is suppressed by FSGG compensation. Fig. \ref{fig:figure4} shows the phase shift across one port of a $20\hbar k$ $^{87}$Rb interferometer with and without FSGG compensation.  To calculate the phase shift as a function of vertical position, we divide each port into four vertical bins and compute the phase shift of each bin relative to the top bin.  FSGG compensation reduces the position-dependent phase shift by a factor of 30, limited by statistical uncertainty.  This method is conceptually similar to the rotation compensation methods of \cite{Gustavson2000, Hogan2009, Lan2012, Sugarbaker2013}, where rotation-induced phase shifts from transverse velocity inhomogeneities are compensated by additional position- and velocity-dependent phase shifts.  We note that methods similar to those employed for rotation compensation can be used to compensate off-axis gravity gradients $T_{xz}$ and $T_{yz}$.  Differential phase offsets from these couplings are currently below our experimental resolution.

The dual-species interferometer demonstrated in this work exhibits unparalleled sensitivity to accelerations.  For a $10\hbar k$ interferometer with $T = 900~\text{ms}$, the differential acceleration sensitivity is $6.4 \times 10^8$ rad/g.  Together with the single-shot differential phase uncertainty of $\sim 40~\text{mrad}$, this implies a relative precision of $\Delta g / g \approx 6 \times 10^{-11}$ per shot or $3 \times 10^{-10}/\sqrt{\text{Hz}}$, which improves on the best published result for a dual-species interferometer by more than three orders of magnitude \cite{Zhou2015}.  Suppressing the relative phase shifts associated with gravity gradients to below $10^{-13}$ is an important step toward allowing this acceleration sensitivity to be utilized for an atom interferometric test of the equivalence principle at an accuracy that is competitive with classical tests \cite{Williams2004, Schlamminger2008}.  We anticipate that planned improvements in the atom source and imaging protocol and a more sophisticated phase extraction algorithm can improve the single-shot phase resolution by an order of magnitude.  In combination with the improvements in initial kinematic control described above, this would pave the way for an atomic equivalence principle test at the $10^{-14}$ level.  FSGG compensation could also be useful for space-based tests of the equivalence principle with even longer interferometer durations \cite{Altschul2015,Williams2016}.  In addition to its application for WEP measurements, the determination of the gravity-gradient-compensating $\Delta f$ could be a valuable tool for precision gravity gradiometry and for measurements of Newton's gravitational constant \cite{Rosi2017}.

\ifdefined\WORDCOUNT
\else
\begin{acknowledgments}
We acknowledge funding from the Defense Threat Reduction Agency, the Jet Propulsion Laboratory, the Office of Naval Research, and the Vannevar Bush Faculty Fellowship program.  We thank Agnetta Cleland and Salvador Gomez for their assistance with this work.
\end{acknowledgments}
\fi

\bibliographystyle{apsrev4-1}

\begin{thebibliography}{36}%
\makeatletter
\providecommand \@ifxundefined [1]{%
 \@ifx{#1\undefined}
}%
\providecommand \@ifnum [1]{%
 \ifnum #1\expandafter \@firstoftwo
 \else \expandafter \@secondoftwo
 \fi
}%
\providecommand \@ifx [1]{%
 \ifx #1\expandafter \@firstoftwo
 \else \expandafter \@secondoftwo
 \fi
}%
\providecommand \natexlab [1]{#1}%
\providecommand \enquote  [1]{``#1''}%
\providecommand \bibnamefont  [1]{#1}%
\providecommand \bibfnamefont [1]{#1}%
\providecommand \citenamefont [1]{#1}%
\providecommand \href@noop [0]{\@secondoftwo}%
\providecommand \href [0]{\begingroup \@sanitize@url \@href}%
\providecommand \@href[1]{\@@startlink{#1}\@@href}%
\providecommand \@@href[1]{\endgroup#1\@@endlink}%
\providecommand \@sanitize@url [0]{\catcode `\\12\catcode `\$12\catcode
  `\&12\catcode `\#12\catcode `\^12\catcode `\_12\catcode `\%12\relax}%
\providecommand \@@startlink[1]{}%
\providecommand \@@endlink[0]{}%
\providecommand \url  [0]{\begingroup\@sanitize@url \@url }%
\providecommand \@url [1]{\endgroup\@href {#1}{\urlprefix }}%
\providecommand \urlprefix  [0]{URL }%
\providecommand \Eprint [0]{\href }%
\providecommand \doibase [0]{http://dx.doi.org/}%
\providecommand \selectlanguage [0]{\@gobble}%
\providecommand \bibinfo  [0]{\@secondoftwo}%
\providecommand \bibfield  [0]{\@secondoftwo}%
\providecommand \translation [1]{[#1]}%
\providecommand \BibitemOpen [0]{}%
\providecommand \bibitemStop [0]{}%
\providecommand \bibitemNoStop [0]{.\EOS\space}%
\providecommand \EOS [0]{\spacefactor3000\relax}%
\providecommand \BibitemShut  [1]{\csname bibitem#1\endcsname}%
\let\auto@bib@innerbib\@empty
\bibitem [{\citenamefont {Williams}\ \emph {et~al.}(2004)\citenamefont
  {Williams}, \citenamefont {Turyshev},\ and\ \citenamefont
  {Boggs}}]{Williams2004}%
  \BibitemOpen
  \bibfield  {author} {\bibinfo {author} {\bibfnamefont {J.~G.}\ \bibnamefont
  {Williams}}, \bibinfo {author} {\bibfnamefont {S.~G.}\ \bibnamefont
  {Turyshev}}, \ and\ \bibinfo {author} {\bibfnamefont {D.~H.}\ \bibnamefont
  {Boggs}},\ }\href {\doibase 10.1103/PhysRevLett.93.261101} {\bibfield
  {journal} {\bibinfo  {journal} {Phys. Rev. Lett.}\ }\textbf {\bibinfo
  {volume} {93}},\ \bibinfo {pages} {261101} (\bibinfo {year}
  {2004})}\BibitemShut {NoStop}%
\bibitem [{\citenamefont {Schlamminger}\ \emph {et~al.}(2008)\citenamefont
  {Schlamminger}, \citenamefont {Choi}, \citenamefont {Wagner}, \citenamefont
  {Gundlach},\ and\ \citenamefont {Adelberger}}]{Schlamminger2008}%
  \BibitemOpen
  \bibfield  {author} {\bibinfo {author} {\bibfnamefont {S.}~\bibnamefont
  {Schlamminger}}, \bibinfo {author} {\bibfnamefont {K.-Y.}\ \bibnamefont
  {Choi}}, \bibinfo {author} {\bibfnamefont {T.~A.}\ \bibnamefont {Wagner}},
  \bibinfo {author} {\bibfnamefont {J.~H.}\ \bibnamefont {Gundlach}}, \ and\
  \bibinfo {author} {\bibfnamefont {E.~G.}\ \bibnamefont {Adelberger}},\ }\href
  {\doibase 10.1103/PhysRevLett.100.041101} {\bibfield  {journal} {\bibinfo
  {journal} {Phys. Rev. Lett.}\ }\textbf {\bibinfo {volume} {100}},\ \bibinfo
  {pages} {041101} (\bibinfo {year} {2008})}\BibitemShut {NoStop}%
\bibitem [{\citenamefont {Fray}\ \emph {et~al.}(2004)\citenamefont {Fray},
  \citenamefont {Diez}, \citenamefont {H\"ansch},\ and\ \citenamefont
  {Weitz}}]{Fray2004}%
  \BibitemOpen
  \bibfield  {author} {\bibinfo {author} {\bibfnamefont {S.}~\bibnamefont
  {Fray}}, \bibinfo {author} {\bibfnamefont {C.~A.}\ \bibnamefont {Diez}},
  \bibinfo {author} {\bibfnamefont {T.~W.}\ \bibnamefont {H\"ansch}}, \ and\
  \bibinfo {author} {\bibfnamefont {M.}~\bibnamefont {Weitz}},\ }\href
  {\doibase 10.1103/PhysRevLett.93.240404} {\bibfield  {journal} {\bibinfo
  {journal} {Phys. Rev. Lett.}\ }\textbf {\bibinfo {volume} {93}},\ \bibinfo
  {pages} {240404} (\bibinfo {year} {2004})}\BibitemShut {NoStop}%
\bibitem [{\citenamefont {Bonnin}\ \emph {et~al.}(2013)\citenamefont {Bonnin},
  \citenamefont {Zahzam}, \citenamefont {Bidel},\ and\ \citenamefont
  {Bresson}}]{Bonnin2013}%
  \BibitemOpen
  \bibfield  {author} {\bibinfo {author} {\bibfnamefont {A.}~\bibnamefont
  {Bonnin}}, \bibinfo {author} {\bibfnamefont {N.}~\bibnamefont {Zahzam}},
  \bibinfo {author} {\bibfnamefont {Y.}~\bibnamefont {Bidel}}, \ and\ \bibinfo
  {author} {\bibfnamefont {A.}~\bibnamefont {Bresson}},\ }\href {\doibase
  10.1103/PhysRevA.88.043615} {\bibfield  {journal} {\bibinfo  {journal} {Phys.
  Rev. A}\ }\textbf {\bibinfo {volume} {88}},\ \bibinfo {pages} {043615}
  (\bibinfo {year} {2013})}\BibitemShut {NoStop}%
\bibitem [{\citenamefont {Schlippert}\ \emph {et~al.}(2014)\citenamefont
  {Schlippert}, \citenamefont {Hartwig}, \citenamefont {Albers}, \citenamefont
  {Richardson}, \citenamefont {Schubert}, \citenamefont {Roura}, \citenamefont
  {Schleich}, \citenamefont {Ertmer},\ and\ \citenamefont
  {Rasel}}]{Schlippert2014}%
  \BibitemOpen
  \bibfield  {author} {\bibinfo {author} {\bibfnamefont {D.}~\bibnamefont
  {Schlippert}}, \bibinfo {author} {\bibfnamefont {J.}~\bibnamefont {Hartwig}},
  \bibinfo {author} {\bibfnamefont {H.}~\bibnamefont {Albers}}, \bibinfo
  {author} {\bibfnamefont {L.~L.}\ \bibnamefont {Richardson}}, \bibinfo
  {author} {\bibfnamefont {C.}~\bibnamefont {Schubert}}, \bibinfo {author}
  {\bibfnamefont {A.}~\bibnamefont {Roura}}, \bibinfo {author} {\bibfnamefont
  {W.~P.}\ \bibnamefont {Schleich}}, \bibinfo {author} {\bibfnamefont
  {W.}~\bibnamefont {Ertmer}}, \ and\ \bibinfo {author} {\bibfnamefont {E.~M.}\
  \bibnamefont {Rasel}},\ }\href {\doibase 10.1103/PhysRevLett.112.203002}
  {\bibfield  {journal} {\bibinfo  {journal} {Phys. Rev. Lett.}\ }\textbf
  {\bibinfo {volume} {112}},\ \bibinfo {pages} {203002} (\bibinfo {year}
  {2014})}\BibitemShut {NoStop}%
\bibitem [{\citenamefont {Tarallo}\ \emph {et~al.}(2014)\citenamefont
  {Tarallo}, \citenamefont {Mazzoni}, \citenamefont {Poli}, \citenamefont
  {Sutyrin}, \citenamefont {Zhang},\ and\ \citenamefont {Tino}}]{Tarallo2014}%
  \BibitemOpen
  \bibfield  {author} {\bibinfo {author} {\bibfnamefont {M.~G.}\ \bibnamefont
  {Tarallo}}, \bibinfo {author} {\bibfnamefont {T.}~\bibnamefont {Mazzoni}},
  \bibinfo {author} {\bibfnamefont {N.}~\bibnamefont {Poli}}, \bibinfo {author}
  {\bibfnamefont {D.~V.}\ \bibnamefont {Sutyrin}}, \bibinfo {author}
  {\bibfnamefont {X.}~\bibnamefont {Zhang}}, \ and\ \bibinfo {author}
  {\bibfnamefont {G.~M.}\ \bibnamefont {Tino}},\ }\href {\doibase
  10.1103/PhysRevLett.113.023005} {\bibfield  {journal} {\bibinfo  {journal}
  {Phys. Rev. Lett.}\ }\textbf {\bibinfo {volume} {113}},\ \bibinfo {pages}
  {023005} (\bibinfo {year} {2014})}\BibitemShut {NoStop}%
\bibitem [{\citenamefont {Zhou}\ \emph {et~al.}(2015)\citenamefont {Zhou},
  \citenamefont {Long}, \citenamefont {Tang}, \citenamefont {Chen},
  \citenamefont {Gao}, \citenamefont {Peng}, \citenamefont {Duan},
  \citenamefont {Zhong}, \citenamefont {Xiong}, \citenamefont {Wang},
  \citenamefont {Zhang},\ and\ \citenamefont {Zhan}}]{Zhou2015}%
  \BibitemOpen
  \bibfield  {author} {\bibinfo {author} {\bibfnamefont {L.}~\bibnamefont
  {Zhou}}, \bibinfo {author} {\bibfnamefont {S.}~\bibnamefont {Long}}, \bibinfo
  {author} {\bibfnamefont {B.}~\bibnamefont {Tang}}, \bibinfo {author}
  {\bibfnamefont {X.}~\bibnamefont {Chen}}, \bibinfo {author} {\bibfnamefont
  {F.}~\bibnamefont {Gao}}, \bibinfo {author} {\bibfnamefont {W.}~\bibnamefont
  {Peng}}, \bibinfo {author} {\bibfnamefont {W.}~\bibnamefont {Duan}}, \bibinfo
  {author} {\bibfnamefont {J.}~\bibnamefont {Zhong}}, \bibinfo {author}
  {\bibfnamefont {Z.}~\bibnamefont {Xiong}}, \bibinfo {author} {\bibfnamefont
  {J.}~\bibnamefont {Wang}}, \bibinfo {author} {\bibfnamefont {Y.}~\bibnamefont
  {Zhang}}, \ and\ \bibinfo {author} {\bibfnamefont {M.}~\bibnamefont {Zhan}},\
  }\href {\doibase 10.1103/PhysRevLett.115.013004} {\bibfield  {journal}
  {\bibinfo  {journal} {Phys. Rev. Lett.}\ }\textbf {\bibinfo {volume} {115}},\
  \bibinfo {pages} {013004} (\bibinfo {year} {2015})}\BibitemShut {NoStop}%
\bibitem [{\citenamefont {Barrett}\ \emph {et~al.}(2016)\citenamefont
  {Barrett}, \citenamefont {Antoni-Micollier}, \citenamefont {Chichet},
  \citenamefont {Battelier}, \citenamefont {L{\'e}v{\`e}que}, \citenamefont
  {Landragin},\ and\ \citenamefont {Bouyer}}]{Barrett2016}%
  \BibitemOpen
  \bibfield  {author} {\bibinfo {author} {\bibfnamefont {B.}~\bibnamefont
  {Barrett}}, \bibinfo {author} {\bibfnamefont {L.}~\bibnamefont
  {Antoni-Micollier}}, \bibinfo {author} {\bibfnamefont {L.}~\bibnamefont
  {Chichet}}, \bibinfo {author} {\bibfnamefont {B.}~\bibnamefont {Battelier}},
  \bibinfo {author} {\bibfnamefont {T.}~\bibnamefont {L{\'e}v{\`e}que}},
  \bibinfo {author} {\bibfnamefont {A.}~\bibnamefont {Landragin}}, \ and\
  \bibinfo {author} {\bibfnamefont {P.}~\bibnamefont {Bouyer}},\ }\href@noop {}
  {\bibfield  {journal} {\bibinfo  {journal} {Nat. Comm.}\ }\textbf {\bibinfo
  {volume} {7}},\ \bibinfo {pages} {13786} (\bibinfo {year}
  {2016})}\BibitemShut {NoStop}%
\bibitem [{\citenamefont {Rosi}\ \emph {et~al.}(2017)\citenamefont {Rosi},
  \citenamefont {D'Amico}, \citenamefont {Cacciapuoti}, \citenamefont
  {Sorrentino}, \citenamefont {Prevedelli}, \citenamefont {Zych}, \citenamefont
  {Brukner},\ and\ \citenamefont {Tino}}]{Rosi2017b}%
  \BibitemOpen
  \bibfield  {author} {\bibinfo {author} {\bibfnamefont {G.}~\bibnamefont
  {Rosi}}, \bibinfo {author} {\bibfnamefont {G.}~\bibnamefont {D`Amico}},
  \bibinfo {author} {\bibfnamefont {L.}~\bibnamefont {Cacciapuoti}}, \bibinfo
  {author} {\bibfnamefont {F.}~\bibnamefont {Sorrentino}}, \bibinfo {author}
  {\bibfnamefont {M.}~\bibnamefont {Prevedelli}}, \bibinfo {author}
  {\bibfnamefont {M.}~\bibnamefont {Zych}}, \bibinfo {author} {\bibfnamefont
  {{\u C}.}~\bibnamefont {Brukner}}, \ and\ \bibinfo {author} {\bibfnamefont
  {G.~M.}\ \bibnamefont {Tino}},\ }\href@noop {} {\bibfield  {journal}
  {\bibinfo  {journal} {Nat. Comm.}\ }\textbf {\bibinfo {volume} {8}},\
  \bibinfo {pages} {15529} (\bibinfo {year} {2017})}\BibitemShut {NoStop}%
\bibitem [{\citenamefont {Berg{\' e}}\ \emph {et~al.}(2015)\citenamefont
  {Berg{\' e}}, \citenamefont {Touboul},\ and\ \citenamefont
  {Rodrigues}}]{Berge2015}%
  \BibitemOpen
  \bibfield  {author} {\bibinfo {author} {\bibfnamefont {J.}~\bibnamefont
  {Berg{\' e}}}, \bibinfo {author} {\bibfnamefont {P.}~\bibnamefont {Touboul}},
  \ and\ \bibinfo {author} {\bibfnamefont {M.}~\bibnamefont {Rodrigues}},\
  }\href {http://stacks.iop.org/1742-6596/610/i=1/a=012009} {\bibfield
  {journal} {\bibinfo  {journal} {Journal of Physics: Conference Series}\
  }\textbf {\bibinfo {volume} {610}},\ \bibinfo {pages} {012009} (\bibinfo
  {year} {2015})}\BibitemShut {NoStop}%
\bibitem [{\citenamefont {Hogan}\ \emph {et~al.}(2009)\citenamefont {Hogan},
  \citenamefont {Johnson},\ and\ \citenamefont {Kasevich}}]{Hogan2009}%
  \BibitemOpen
  \bibfield  {author} {\bibinfo {author} {\bibfnamefont {J.~M.}\ \bibnamefont
  {Hogan}}, \bibinfo {author} {\bibfnamefont {D.~M.~S.}\ \bibnamefont
  {Johnson}}, \ and\ \bibinfo {author} {\bibfnamefont {M.~A.}\ \bibnamefont
  {Kasevich}},\ }in\ \href {http://arxiv.org/abs/0806.3261} {\emph {\bibinfo
  {booktitle} {Proc. Int. Sch. Phys. ``Enrico Fermi'' Atom Opt. Sp. Phys.}}},\
  \bibinfo {editor} {edited by\ \bibinfo {editor} {\bibfnamefont
  {E.}~\bibnamefont {Arimondo}}, \bibinfo {editor} {\bibfnamefont
  {W.}~\bibnamefont {Ertmer}}, \ and\ \bibinfo {editor} {\bibfnamefont {W.~P.}\
  \bibnamefont {Schleich}}}\ (\bibinfo  {publisher} {IOS Press},\ \bibinfo
  {address} {Amsterdam},\ \bibinfo {year} {2009})\ pp.\ \bibinfo {pages}
  {411--447},\ \Eprint {http://arxiv.org/abs/0806.3261} {arXiv:0806.3261}
  \BibitemShut {NoStop}%
\bibitem [{\citenamefont {Altschul}\ \emph {et~al.}(2015)\citenamefont
  {Altschul}, \citenamefont {Bailey}, \citenamefont {Blanchet}, \citenamefont
  {Bongs}, \citenamefont {Bouyer}, \citenamefont {Cacciapuoti}, \citenamefont
  {Capozziello}, \citenamefont {Gaaloul}, \citenamefont {Giulini},
  \citenamefont {Hartwig}, \citenamefont {Iess}, \citenamefont {Jetzer},
  \citenamefont {Landragin}, \citenamefont {Rasel}, \citenamefont {Reynaud},
  \citenamefont {Schiller}, \citenamefont {Schubert}, \citenamefont
  {Sorrentino}, \citenamefont {Sterr}, \citenamefont {Tasson}, \citenamefont
  {Tino}, \citenamefont {Tuckey},\ and\ \citenamefont {Wolf}}]{Altschul2015}%
  \BibitemOpen
  \bibfield  {author} {\bibinfo {author} {\bibfnamefont {B.}~\bibnamefont
  {Altschul}}, \bibinfo {author} {\bibfnamefont {Q.~G.}\ \bibnamefont
  {Bailey}}, \bibinfo {author} {\bibfnamefont {L.}~\bibnamefont {Blanchet}},
  \bibinfo {author} {\bibfnamefont {K.}~\bibnamefont {Bongs}}, \bibinfo
  {author} {\bibfnamefont {P.}~\bibnamefont {Bouyer}}, \bibinfo {author}
  {\bibfnamefont {L.}~\bibnamefont {Cacciapuoti}}, \bibinfo {author}
  {\bibfnamefont {S.}~\bibnamefont {Capozziello}}, \bibinfo {author}
  {\bibfnamefont {N.}~\bibnamefont {Gaaloul}}, \bibinfo {author} {\bibfnamefont
  {D.}~\bibnamefont {Giulini}}, \bibinfo {author} {\bibfnamefont
  {J.}~\bibnamefont {Hartwig}}, \bibinfo {author} {\bibfnamefont
  {L.}~\bibnamefont {Iess}}, \bibinfo {author} {\bibfnamefont {P.}~\bibnamefont
  {Jetzer}}, \bibinfo {author} {\bibfnamefont {A.}~\bibnamefont {Landragin}},
  \bibinfo {author} {\bibfnamefont {E.}~\bibnamefont {Rasel}}, \bibinfo
  {author} {\bibfnamefont {S.}~\bibnamefont {Reynaud}}, \bibinfo {author}
  {\bibfnamefont {S.}~\bibnamefont {Schiller}}, \bibinfo {author}
  {\bibfnamefont {C.}~\bibnamefont {Schubert}}, \bibinfo {author}
  {\bibfnamefont {F.}~\bibnamefont {Sorrentino}}, \bibinfo {author}
  {\bibfnamefont {U.}~\bibnamefont {Sterr}}, \bibinfo {author} {\bibfnamefont
  {J.~D.}\ \bibnamefont {Tasson}}, \bibinfo {author} {\bibfnamefont {G.~M.}\
  \bibnamefont {Tino}}, \bibinfo {author} {\bibfnamefont {P.}~\bibnamefont
  {Tuckey}}, \ and\ \bibinfo {author} {\bibfnamefont {P.}~\bibnamefont
  {Wolf}},\ }\href {\doibase https://doi.org/10.1016/j.asr.2014.07.014}
  {\bibfield  {journal} {\bibinfo  {journal} {Advances in Space Research}\
  }\textbf {\bibinfo {volume} {55}},\ \bibinfo {pages} {501 } (\bibinfo {year}
  {2015})}\BibitemShut {NoStop}%
\bibitem [{\citenamefont {Hartwig}\ \citenamefont
  {et al}(2015)}]{Hartwig2015}%
  \BibitemOpen
  \bibfield  {author} {\bibinfo {author} {\bibfnamefont {J.}~\bibnamefont
  {Hartwig}}\ \textit{et al.},\ }\href {\doibase
  10.1088/1367-2630/17/3/035011} {\bibfield  {journal} {\bibinfo  {journal}
  {New J. Phys.}\ }\textbf {\bibinfo {volume} {17}},\ \bibinfo {pages} {35011}
  (\bibinfo {year} {2015})}\BibitemShut {NoStop}%
\bibitem [{\citenamefont {Williams}\ \emph {et~al.}(2016)\citenamefont
  {Williams}, \citenamefont {Chiow}, \citenamefont {Yu},\ and\ \citenamefont
  {M{\"u}ller}}]{Williams2016}%
  \BibitemOpen
  \bibfield  {author} {\bibinfo {author} {\bibfnamefont {J.}~\bibnamefont
  {Williams}}, \bibinfo {author} {\bibfnamefont {S.-w.}\ \bibnamefont {Chiow}},
  \bibinfo {author} {\bibfnamefont {N.}~\bibnamefont {Yu}}, \ and\ \bibinfo
  {author} {\bibfnamefont {H.}~\bibnamefont {M{\"u}ller}},\ }\href
  {http://stacks.iop.org/1367-2630/18/i=2/a=025018} {\bibfield  {journal}
  {\bibinfo  {journal} {New J. Phys.}\ }\textbf {\bibinfo {volume} {18}},\
  \bibinfo {pages} {025018} (\bibinfo {year} {2016})}\BibitemShut {NoStop}%
\bibitem [{\citenamefont {Carroll}(2004)}]{Carroll2004a}%
  \BibitemOpen
  \bibfield  {author} {\bibinfo {author} {\bibfnamefont {S.~M.}\ \bibnamefont
  {Carroll}},\ }\href@noop {} {\emph {\bibinfo {title} {{Spacetime and
  Geometry: An Introduction to General Relativity}}}}\ (\bibinfo  {publisher}
  {Addison-Wesley},\ \bibinfo {address} {San Francisco},\ \bibinfo {year}
  {2004})\BibitemShut {NoStop}%
\bibitem [{\citenamefont {Dimopoulos}\ \emph {et~al.}(2008)\citenamefont
  {Dimopoulos}, \citenamefont {Graham}, \citenamefont {Hogan},\ and\
  \citenamefont {Kasevich}}]{Dimopoulos2008a}%
  \BibitemOpen
  \bibfield  {author} {\bibinfo {author} {\bibfnamefont {S.}~\bibnamefont
  {Dimopoulos}}, \bibinfo {author} {\bibfnamefont {P.}~\bibnamefont {Graham}},
  \bibinfo {author} {\bibfnamefont {J.}~\bibnamefont {Hogan}}, \ and\ \bibinfo
  {author} {\bibfnamefont {M.}~\bibnamefont {Kasevich}},\ }\href {\doibase
  10.1103/PhysRevD.78.042003} {\bibfield  {journal} {\bibinfo  {journal} {Phys.
  Rev. D}\ }\textbf {\bibinfo {volume} {78}},\ \bibinfo {pages} {42003}
  (\bibinfo {year} {2008})}\BibitemShut {NoStop}%
\bibitem [{\citenamefont {Graham}\ \emph {et~al.}(2016)\citenamefont {Graham},
  \citenamefont {Kaplan}, \citenamefont {Mardon}, \citenamefont {Rajendran},\
  and\ \citenamefont {Terrano}}]{Graham2016}%
  \BibitemOpen
  \bibfield  {author} {\bibinfo {author} {\bibfnamefont {P.~W.}\ \bibnamefont
  {Graham}}, \bibinfo {author} {\bibfnamefont {D.~E.}\ \bibnamefont {Kaplan}},
  \bibinfo {author} {\bibfnamefont {J.}~\bibnamefont {Mardon}}, \bibinfo
  {author} {\bibfnamefont {S.}~\bibnamefont {Rajendran}}, \ and\ \bibinfo
  {author} {\bibfnamefont {W.~A.}\ \bibnamefont {Terrano}},\ }\href {\doibase
  10.1103/PhysRevD.93.075029} {\bibfield  {journal} {\bibinfo  {journal} {Phys.
  Rev. D}\ }\textbf {\bibinfo {volume} {93}},\ \bibinfo {pages} {075029}
  (\bibinfo {year} {2016})}\BibitemShut {NoStop}%
\bibitem [{\citenamefont {Nobili}(2016)}]{Nobili2016}%
  \BibitemOpen
  \bibfield  {author} {\bibinfo {author} {\bibfnamefont {A.~M.}\ \bibnamefont
  {Nobili}},\ }\href {\doibase 10.1103/PhysRevA.93.023617} {\bibfield
  {journal} {\bibinfo  {journal} {Phys. Rev. A}\ }\textbf {\bibinfo {volume}
  {93}},\ \bibinfo {pages} {023617} (\bibinfo {year} {2016})}\BibitemShut
  {NoStop}%
\bibitem [{\citenamefont {Roura}(2017)}]{Roura2017}%
  \BibitemOpen
  \bibfield  {author} {\bibinfo {author} {\bibfnamefont {A.}~\bibnamefont
  {Roura}},\ }\href {\doibase 10.1103/PhysRevLett.118.160401} {\bibfield
  {journal} {\bibinfo  {journal} {Phys. Rev. Lett.}\ }\textbf {\bibinfo
  {volume} {118}},\ \bibinfo {pages} {160401} (\bibinfo {year}
  {2017})}\BibitemShut {NoStop}%
\bibitem [{\citenamefont {Bord\'{e}}(1989)}]{Borde1989}%
  \BibitemOpen
  \bibfield  {author} {\bibinfo {author} {\bibfnamefont {C.~J.}\ \bibnamefont
  {Bord\'{e}}},\ }\href {\doibase
  http://dx.doi.org/10.1016/0375-9601(89)90537-9} {\bibfield  {journal}
  {\bibinfo  {journal} {Phys. Lett. A}\ }\textbf {\bibinfo {volume} {140}},\
  \bibinfo {pages} {10} (\bibinfo {year} {1989})}\BibitemShut {NoStop}%
\bibitem [{\citenamefont {Kasevich}\ and\ \citenamefont
  {Chu}(1991)}]{Kasevich1991}%
  \BibitemOpen
  \bibfield  {author} {\bibinfo {author} {\bibfnamefont {M.}~\bibnamefont
  {Kasevich}}\ and\ \bibinfo {author} {\bibfnamefont {S.}~\bibnamefont {Chu}},\
  }\href {\doibase 10.1103/PhysRevLett.67.181} {\bibfield  {journal} {\bibinfo
  {journal} {Phys. Rev. Lett.}\ }\textbf {\bibinfo {volume} {67}},\ \bibinfo
  {pages} {181} (\bibinfo {year} {1991})}\BibitemShut {NoStop}%
\bibitem [{\citenamefont {Biedermann}\ \emph {et~al.}(2015)\citenamefont
  {Biedermann}, \citenamefont {Wu}, \citenamefont {Deslauriers}, \citenamefont
  {Roy}, \citenamefont {Mahadeswaraswamy},\ and\ \citenamefont
  {Kasevich}}]{Biedermann2015}%
  \BibitemOpen
  \bibfield  {author} {\bibinfo {author} {\bibfnamefont {G.~W.}\ \bibnamefont
  {Biedermann}}, \bibinfo {author} {\bibfnamefont {X.}~\bibnamefont {Wu}},
  \bibinfo {author} {\bibfnamefont {L.}~\bibnamefont {Deslauriers}}, \bibinfo
  {author} {\bibfnamefont {S.}~\bibnamefont {Roy}}, \bibinfo {author}
  {\bibfnamefont {C.}~\bibnamefont {Mahadeswaraswamy}}, \ and\ \bibinfo
  {author} {\bibfnamefont {M.~A.}\ \bibnamefont {Kasevich}},\ }\href {\doibase
  10.1103/PhysRevA.91.033629} {\bibfield  {journal} {\bibinfo  {journal} {Phys.
  Rev. A}\ }\textbf {\bibinfo {volume} {91}},\ \bibinfo {pages} {033629}
  (\bibinfo {year} {2015})}\BibitemShut {NoStop}%
\bibitem [{\citenamefont {Dickerson}\ \emph {et~al.}(2013)\citenamefont
  {Dickerson}, \citenamefont {Hogan}, \citenamefont {Sugarbaker}, \citenamefont
  {Johnson},\ and\ \citenamefont {Kasevich}}]{Dickerson2013}%
  \BibitemOpen
  \bibfield  {author} {\bibinfo {author} {\bibfnamefont {S.~M.}\ \bibnamefont
  {Dickerson}}, \bibinfo {author} {\bibfnamefont {J.~M.}\ \bibnamefont
  {Hogan}}, \bibinfo {author} {\bibfnamefont {A.}~\bibnamefont {Sugarbaker}},
  \bibinfo {author} {\bibfnamefont {D.~M.~S.}\ \bibnamefont {Johnson}}, \ and\
  \bibinfo {author} {\bibfnamefont {M.~A.}\ \bibnamefont {Kasevich}},\ }\href
  {\doibase 10.1103/PhysRevLett.111.083001} {\bibfield  {journal} {\bibinfo
  {journal} {Phys. Rev. Lett.}\ }\textbf {\bibinfo {volume} {111}},\ \bibinfo
  {pages} {83001} (\bibinfo {year} {2013})}\BibitemShut {NoStop}%
\bibitem [{\citenamefont {Sugarbaker}\ \emph {et~al.}(2013)\citenamefont
  {Sugarbaker}, \citenamefont {Dickerson}, \citenamefont {Hogan}, \citenamefont
  {Johnson},\ and\ \citenamefont {Kasevich}}]{Sugarbaker2013}%
  \BibitemOpen
  \bibfield  {author} {\bibinfo {author} {\bibfnamefont {A.}~\bibnamefont
  {Sugarbaker}}, \bibinfo {author} {\bibfnamefont {S.~M.}\ \bibnamefont
  {Dickerson}}, \bibinfo {author} {\bibfnamefont {J.~M.}\ \bibnamefont
  {Hogan}}, \bibinfo {author} {\bibfnamefont {D.~M.~S.}\ \bibnamefont
  {Johnson}}, \ and\ \bibinfo {author} {\bibfnamefont {M.~A.}\ \bibnamefont
  {Kasevich}},\ }\href {\doibase 10.1103/PhysRevLett.111.113002} {\bibfield
  {journal} {\bibinfo  {journal} {Phys. Rev. Lett.}\ }\textbf {\bibinfo
  {volume} {111}},\ \bibinfo {pages} {113002} (\bibinfo {year}
  {2013})}\BibitemShut {NoStop}%
\bibitem [{\citenamefont {Kovachy}\ \emph
  {et~al.}(2015{\natexlab{a}})\citenamefont {Kovachy}, \citenamefont {Hogan},
  \citenamefont {Sugarbaker}, \citenamefont {Dickerson}, \citenamefont
  {Donnelly}, \citenamefont {Overstreet},\ and\ \citenamefont
  {Kasevich}}]{Kovachy2015}%
  \BibitemOpen
  \bibfield  {author} {\bibinfo {author} {\bibfnamefont {T.}~\bibnamefont
  {Kovachy}}, \bibinfo {author} {\bibfnamefont {J.~M.}\ \bibnamefont {Hogan}},
  \bibinfo {author} {\bibfnamefont {A.}~\bibnamefont {Sugarbaker}}, \bibinfo
  {author} {\bibfnamefont {S.~M.}\ \bibnamefont {Dickerson}}, \bibinfo {author}
  {\bibfnamefont {C.~A.}\ \bibnamefont {Donnelly}}, \bibinfo {author}
  {\bibfnamefont {C.}~\bibnamefont {Overstreet}}, \ and\ \bibinfo {author}
  {\bibfnamefont {M.~A.}\ \bibnamefont {Kasevich}},\ }\href {\doibase
  10.1103/PhysRevLett.114.143004} {\bibfield  {journal} {\bibinfo  {journal}
  {Phys. Rev. Lett.}\ }\textbf {\bibinfo {volume} {114}},\ \bibinfo {pages}
  {143004} (\bibinfo {year} {2015}{\natexlab{a}})}\BibitemShut {NoStop}%
\bibitem [{\citenamefont {Kovachy}\ \emph
  {et~al.}(2015{\natexlab{b}})\citenamefont {Kovachy}, \citenamefont
  {Asenbaum}, \citenamefont {Overstreet}, \citenamefont {Donnelly},
  \citenamefont {Dickerson}, \citenamefont {Sugarbaker}, \citenamefont
  {Hogan},\ and\ \citenamefont {Kasevich}}]{Kovachy2015a}%
  \BibitemOpen
  \bibfield  {author} {\bibinfo {author} {\bibfnamefont {T.}~\bibnamefont
  {Kovachy}}, \bibinfo {author} {\bibfnamefont {P.}~\bibnamefont {Asenbaum}},
  \bibinfo {author} {\bibfnamefont {C.}~\bibnamefont {Overstreet}}, \bibinfo
  {author} {\bibfnamefont {C.~A.}\ \bibnamefont {Donnelly}}, \bibinfo {author}
  {\bibfnamefont {S.~M.}\ \bibnamefont {Dickerson}}, \bibinfo {author}
  {\bibfnamefont {A.}~\bibnamefont {Sugarbaker}}, \bibinfo {author}
  {\bibfnamefont {J.~M.}\ \bibnamefont {Hogan}}, \ and\ \bibinfo {author}
  {\bibfnamefont {M.~A.}\ \bibnamefont {Kasevich}},\ }\href {\doibase
  10.1038/nature16155} {\bibfield  {journal} {\bibinfo  {journal} {Nature}\
  }\textbf {\bibinfo {volume} {528}},\ \bibinfo {pages} {530} (\bibinfo {year}
  {2015}{\natexlab{b}})}\BibitemShut {NoStop}%
\bibitem [{\citenamefont {Asenbaum}\ \emph {et~al.}(2017)\citenamefont
  {Asenbaum}, \citenamefont {Overstreet}, \citenamefont {Kovachy},
  \citenamefont {Brown}, \citenamefont {Hogan},\ and\ \citenamefont
  {Kasevich}}]{Asenbaum2017}%
  \BibitemOpen
  \bibfield  {author} {\bibinfo {author} {\bibfnamefont {P.}~\bibnamefont
  {Asenbaum}}, \bibinfo {author} {\bibfnamefont {C.}~\bibnamefont
  {Overstreet}}, \bibinfo {author} {\bibfnamefont {T.}~\bibnamefont {Kovachy}},
  \bibinfo {author} {\bibfnamefont {D.~D.}\ \bibnamefont {Brown}}, \bibinfo
  {author} {\bibfnamefont {J.~M.}\ \bibnamefont {Hogan}}, \ and\ \bibinfo
  {author} {\bibfnamefont {M.~A.}\ \bibnamefont {Kasevich}},\ }\href {\doibase
  10.1103/PhysRevLett.118.183602} {\bibfield  {journal} {\bibinfo  {journal}
  {Phys. Rev. Lett.}\ }\textbf {\bibinfo {volume} {118}},\ \bibinfo {pages}
  {183602} (\bibinfo {year} {2017})}\BibitemShut {NoStop}%
\bibitem [{\citenamefont {Ahlers}\ \emph {et~al.}(2016)\citenamefont {Ahlers},
  \citenamefont {M\"{u}ntinga}, \citenamefont {Wenzlawski}, \citenamefont
  {Krutzik}, \citenamefont {Tackmann}, \citenamefont {Abend}, \citenamefont
  {Gaaloul}, \citenamefont {Giese}, \citenamefont {Roura}, \citenamefont
  {Kuhl}, \citenamefont {L\"{a}mmerzahl}, \citenamefont {Peters}, \citenamefont
  {Windpassinger}, \citenamefont {Sengstock}, \citenamefont {Schleich},
  \citenamefont {Ertmer},\ and\ \citenamefont {Rasel}}]{Ahlers2016}%
  \BibitemOpen
  \bibfield  {author} {\bibinfo {author} {\bibfnamefont {H.}~\bibnamefont
  {Ahlers}}, \bibinfo {author} {\bibfnamefont {H.}~\bibnamefont
  {M\"{u}ntinga}}, \bibinfo {author} {\bibfnamefont {A.}~\bibnamefont
  {Wenzlawski}}, \bibinfo {author} {\bibfnamefont {M.}~\bibnamefont {Krutzik}},
  \bibinfo {author} {\bibfnamefont {G.}~\bibnamefont {Tackmann}}, \bibinfo
  {author} {\bibfnamefont {S.}~\bibnamefont {Abend}}, \bibinfo {author}
  {\bibfnamefont {N.}~\bibnamefont {Gaaloul}}, \bibinfo {author} {\bibfnamefont
  {E.}~\bibnamefont {Giese}}, \bibinfo {author} {\bibfnamefont
  {A.}~\bibnamefont {Roura}}, \bibinfo {author} {\bibfnamefont
  {R.}~\bibnamefont {Kuhl}}, \bibinfo {author} {\bibfnamefont {C.}~\bibnamefont
  {L\"{a}mmerzahl}}, \bibinfo {author} {\bibfnamefont {A.}~\bibnamefont
  {Peters}}, \bibinfo {author} {\bibfnamefont {P.}~\bibnamefont
  {Windpassinger}}, \bibinfo {author} {\bibfnamefont {K.}~\bibnamefont
  {Sengstock}}, \bibinfo {author} {\bibfnamefont {W.~P.}\ \bibnamefont
  {Schleich}}, \bibinfo {author} {\bibfnamefont {W.}~\bibnamefont {Ertmer}}, \
  and\ \bibinfo {author} {\bibfnamefont {E.~M.}\ \bibnamefont {Rasel}},\ }\href
  {\doibase 10.1103/PhysRevLett.116.173601} {\bibfield  {journal} {\bibinfo
  {journal} {Phys. Rev. Lett.}\ }\textbf {\bibinfo {volume} {116}},\ \bibinfo
  {pages} {173601} (\bibinfo {year} {2016})}\BibitemShut {NoStop}%
\bibitem [{\citenamefont {Boyd}\ \emph {et~al.}(2007)\citenamefont {Boyd},
  \citenamefont {Zelevinsky}, \citenamefont {Ludlow}, \citenamefont {Blatt},
  \citenamefont {Zanon-Willette}, \citenamefont {Foreman},\ and\ \citenamefont
  {Ye}}]{Boyd2007}%
  \BibitemOpen
  \bibfield  {author} {\bibinfo {author} {\bibfnamefont {M.~M.}\ \bibnamefont
  {Boyd}}, \bibinfo {author} {\bibfnamefont {T.}~\bibnamefont {Zelevinsky}},
  \bibinfo {author} {\bibfnamefont {A.~D.}\ \bibnamefont {Ludlow}}, \bibinfo
  {author} {\bibfnamefont {S.}~\bibnamefont {Blatt}}, \bibinfo {author}
  {\bibfnamefont {T.}~\bibnamefont {Zanon-Willette}}, \bibinfo {author}
  {\bibfnamefont {S.~M.}\ \bibnamefont {Foreman}}, \ and\ \bibinfo {author}
  {\bibfnamefont {J.}~\bibnamefont {Ye}},\ }\href {\doibase
  10.1103/PhysRevA.76.022510} {\bibfield  {journal} {\bibinfo  {journal} {Phys.
  Rev. A}\ }\textbf {\bibinfo {volume} {76}},\ \bibinfo {pages} {022510}
  (\bibinfo {year} {2007})}\BibitemShut {NoStop}%
\bibitem [{\citenamefont {Ye}\ \emph {et~al.}(2008)\citenamefont {Ye},
  \citenamefont {Kimble},\ and\ \citenamefont {Katori}}]{Ye2008}%
  \BibitemOpen
  \bibfield  {author} {\bibinfo {author} {\bibfnamefont {J.}~\bibnamefont
  {Ye}}, \bibinfo {author} {\bibfnamefont {H.~J.}\ \bibnamefont {Kimble}}, \
  and\ \bibinfo {author} {\bibfnamefont {H.}~\bibnamefont {Katori}},\ }\href
  {\doibase 10.1126/science.1148259} {\bibfield  {journal} {\bibinfo  {journal}
  {Science}\ }\textbf {\bibinfo {volume} {320}},\ \bibinfo {pages} {1734}
  (\bibinfo {year} {2008})}\BibitemShut {NoStop}%
\bibitem [{\citenamefont {Kasevich}\ \emph {et~al.}(1991)\citenamefont
  {Kasevich}, \citenamefont {Weiss}, \citenamefont {Riis}, \citenamefont
  {Moler}, \citenamefont {Kasapi},\ and\ \citenamefont {Chu}}]{Kasevich1991b}%
  \BibitemOpen
  \bibfield  {author} {\bibinfo {author} {\bibfnamefont {M.}~\bibnamefont
  {Kasevich}}, \bibinfo {author} {\bibfnamefont {D.~S.}\ \bibnamefont {Weiss}},
  \bibinfo {author} {\bibfnamefont {E.}~\bibnamefont {Riis}}, \bibinfo {author}
  {\bibfnamefont {K.}~\bibnamefont {Moler}}, \bibinfo {author} {\bibfnamefont
  {S.}~\bibnamefont {Kasapi}}, \ and\ \bibinfo {author} {\bibfnamefont
  {S.}~\bibnamefont {Chu}},\ }\href {\doibase 10.1103/PhysRevLett.66.2297}
  {\bibfield  {journal} {\bibinfo  {journal} {Phys. Rev. Lett.}\ }\textbf
  {\bibinfo {volume} {66}},\ \bibinfo {pages} {2297} (\bibinfo {year}
  {1991})}\BibitemShut {NoStop}%
\bibitem [{\citenamefont {Steck}(2008{\natexlab{a}})}]{Steck85}%
  \BibitemOpen
  \bibfield  {author} {\bibinfo {author} {\bibfnamefont {D.~A.}\ \bibnamefont
  {Steck}},\ }\href {http://steck.us/alkalidata/rubidium85numbers.pdf} {\emph
  {\bibinfo {title} {{Rubidium 85 D Line Data}}}},\ \bibinfo {type} {Tech.
  Rep.}\ (\bibinfo {year} {2008})\BibitemShut {NoStop}%
\bibitem [{\citenamefont {Steck}(2008{\natexlab{b}})}]{Steck87}%
  \BibitemOpen
  \bibfield  {author} {\bibinfo {author} {\bibfnamefont {D.~A.}\ \bibnamefont
  {Steck}},\ }\href {http://steck.us/alkalidata/rubidium87numbers.1.6.pdf}
  {\emph {\bibinfo {title} {{Rubidium 87 D Line Data}}}},\ \bibinfo {type}
  {Tech. Rep.}\ (\bibinfo {year} {2008})\BibitemShut {NoStop}%
\bibitem [{\citenamefont {Gustavson}\ \emph {et~al.}(2000)\citenamefont
  {Gustavson}, \citenamefont {Landragin},\ and\ \citenamefont
  {Kasevich}}]{Gustavson2000}%
  \BibitemOpen
  \bibfield  {author} {\bibinfo {author} {\bibfnamefont {T.~L.}\ \bibnamefont
  {Gustavson}}, \bibinfo {author} {\bibfnamefont {A.}~\bibnamefont
  {Landragin}}, \ and\ \bibinfo {author} {\bibfnamefont {M.~A.}\ \bibnamefont
  {Kasevich}},\ }\href {http://stacks.iop.org/0264-9381/17/i=12/a=311}
  {\bibfield  {journal} {\bibinfo  {journal} {Classical and Quantum Gravity}\
  }\textbf {\bibinfo {volume} {17}},\ \bibinfo {pages} {2385} (\bibinfo {year}
  {2000})}\BibitemShut {NoStop}%
\bibitem [{\citenamefont {Lan}\ \emph {et~al.}(2012)\citenamefont {Lan},
  \citenamefont {Kuan}, \citenamefont {Estey}, \citenamefont {Haslinger},\ and\
  \citenamefont {M\"uller}}]{Lan2012}%
  \BibitemOpen
  \bibfield  {author} {\bibinfo {author} {\bibfnamefont {S.-Y.}\ \bibnamefont
  {Lan}}, \bibinfo {author} {\bibfnamefont {P.-C.}\ \bibnamefont {Kuan}},
  \bibinfo {author} {\bibfnamefont {B.}~\bibnamefont {Estey}}, \bibinfo
  {author} {\bibfnamefont {P.}~\bibnamefont {Haslinger}}, \ and\ \bibinfo
  {author} {\bibfnamefont {H.}~\bibnamefont {M\"uller}},\ }\href {\doibase
  10.1103/PhysRevLett.108.090402} {\bibfield  {journal} {\bibinfo  {journal}
  {Phys. Rev. Lett.}\ }\textbf {\bibinfo {volume} {108}},\ \bibinfo {pages}
  {090402} (\bibinfo {year} {2012})}\BibitemShut {NoStop}%
\bibitem [{\citenamefont {Rosi}(2017)}]{Rosi2017}%
  \BibitemOpen
  \bibfield  {author} {\bibinfo {author} {\bibfnamefont {G.}~\bibnamefont
  {Rosi}},\ }\href {http://iopscience.iop.org/10.1088/1681-7575/aa8fd8}
  {\bibfield  {journal} {\bibinfo  {journal} {Metrologia (manuscript accepted,
  early posting)}\ } (\bibinfo {year} {2017})}\BibitemShut {NoStop}%
\end{thebibliography}
%

\ifdefined\WORDCOUNT
    \end{document}
\else
\fi

\end{document}